\documentclass{article}
\usepackage{spconf,amsmath,graphicx}
\usepackage{enumitem}
\setlist{nosep, leftmargin=14pt}

\usepackage{mwe} 
\usepackage{multirow}
\usepackage{makecell}

\title{Zero-shot Bias Correction: Efficient MR Image Inhomogeneity Reduction Without Any Data}
\name{Hongxu Yang$^{\star}$ , Edina Timko$^{\star}$, Brice Fernandez$^{\dagger}$ \thanks{Contact author: hongxu.yang@gehealthcare.com}}
\address{$^{\star}$STO-Artificial Intelligence \& Machine Learning, GE HealthCare\\
$^{\dagger}$ IMG-MR-Applications \& Workflow, GE HealthCare}
\begin{document}
%
\maketitle
\begin{abstract}
In recent years, deep neural networks for image inhomogeneity reduction have shown promising results. However, current methods with (un)supervised solutions require preparing a training dataset, which is expensive and laborious for data collection. In this work, we demonstrate a novel zero-shot deep neural networks, which requires no data for pre-training and dedicated assumption of the bias field. The designed light-weight CNN enables an efficient zero-shot adaptation for bias-corrupted image correction. Our method provides a novel solution to mitigate the biased corrupted image as iterative homogeneity refinement, which therefore ensures the considered issue can be solved easier with stable convergence of zero-shot optimization. Extensive comparison on different datasets show that the proposed method performs better than current data-free N4 methods in both efficiency and accuracy. 
\end{abstract}
\section{Introduction}
\label{sec:intro}
Magnetic Resonance Imaging (MRI) provides detailed anatomy information of patient, which are essential for accurate and precise diagnosis and analysis. Nevertheless, MRI images are commonly degraded by several different artifacts, which hampers the information in the obtained data. Image inhomogeneity in MRI, which is commonly caused by bias field from imperfect scanner (such as phased-array coils) and magnetic filed distribution. With this scanning phenomenon in MRI, the signal intensity within the same anatomy would have a larger low-frequency variance across the obtained volume. Although this inhomogeneity might not be impacting the diagnoses by experienced clinicians, it would drastically degrade the subsequent image processing algorithms and methods, such as registration and segmentation in the fully automated image pipeline. Inhomogeneity removal in MRI has been studied during past decades, and the approaches are categorized as prospective and retrospective methods~\cite{vovk2007review}. The prospective methods are commonly correcting the bias in the scanning step, which requires dedicated hardware and software design, and therefore economically expensive for medical device companies. In contrast, retrospective methods are processing the obtained images by various image processing techniques to reduce the inhomogeneity and bias in the data. The commonly considered correction methods can be classified into non-learning-based method and learning-based method. Specifically, the most popular non-learning-based method of bias correction is the nonparametric nonuniformity intensity normalization (N3) method~\cite{sled1998nonparametric}, and also its modification N4~\cite{tustison2010n4itk}. Although these methods mostly achieved promising performances and widely accepted by communities of MR imaging, iteration computing on histogram and frequency correction has made these methods time consuming~\cite{perezunsupervised}. In addition, as shown in the literature, the N4 method tends to produce unrealistic results when the variance of the bias values are too large~\cite{liang2023unsupervised}. Recent years, (deep) learning-based correction have received remarkable attention in different artifacts reduction application. Supervised deep learning methods were developed for bias-free MRI generation with great success~\cite{goldfryd2021deep, chen2021abcnet}. However, as known in the community, supervised learning methods are suffering lack of large amount of training images with ground truth. To tackle this limitation, unsupervised inhomogeneity correction methods with deep learning (DL) strategies were proposed, such as implicitly trained CNN~\cite{simko2022mri} and histogram-based entropy correction method~\cite{perezunsupervised}. The aforementioned methods achieved comparable performances w.r.t. commonly applied N4 algorithm while reducing the annotation efforts of ground truth generation. Nevertheless, there are still several considerations about these methods. First, the implicitly trained CNN requires proper bias field simulation of the images, which is highly correlated to devices and anatomies. Therefore, dedicated research and training data preparation is inevitable. Second, to train a model that can be generalized to different situation, laborious data collection is needed and can be expensive for data preparations. Third, the network trained on such dataset suffers performance drop if the testing datasets are derived from different data distributions. All the above concerns motivated us to study the data-free method.

\begin{figure*}[ht!]
\centering{\includegraphics[width=17cm]{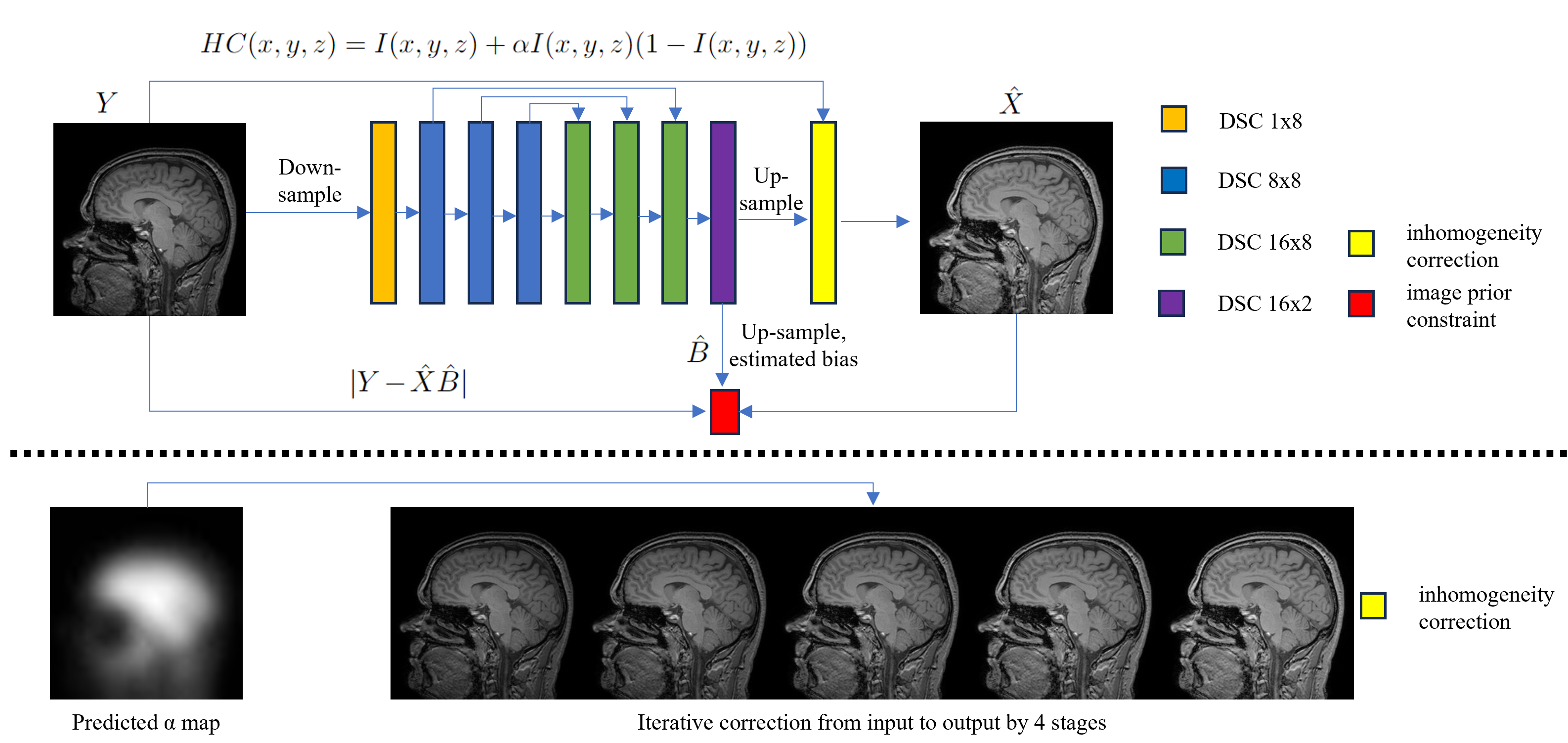}}
\caption{Top: The light-weight zero-shot inhomogeneity correction model. The input 3D volumetric data is processed by feature extraction blocks to generate parametric maps for iterative homogeneity refinement and image prior loss. DSC: Depthwise Separable Convolutions. The proposed model has around 3k trainable parameters. Bottom: Iterative inhomogeneity correction with the predicted $\alpha$ map per pixel.}
\label{model_design}
\end{figure*}

This paper presents a zero-shot inhomogeneity correction method based on deep learning. The method, as will be explained in the later section, composites two functions of homogeneity refinement and image prior constraint modules. Our solution is able to correct the bias corrupted images without any training data, by simply performing few steps of iteration at testing time with light-weight Convolutional Neural Network (CNN). The novel zero-shot method has three major contributions. First, common data-free methods requires solving an optimization problem at inference time. Standard CNN with huge number of parameter count would lead to large memory cost in hardware, e.g. GPU RAM and long processing time. In this work, we utilize a simple CNN architecture with only around 3k parameters, therefore the computational cost is drastically reduced within seconds. Second, our proposal provides a solution of bias field correction by iterative inhomogeneity refinement, therefore the considered problem is easier to be solved for zero-shot situation. Third, the extensive validation on public brain datasets and private pelvis datasets show the presented zero-shot method is comparable or even better than N4 method in both efficiency and accuracy performances. 

\section{Method}
The proposed zero-shot inhomogeneity correction method is depicted in Fig.~\ref{model_design}. The considered CNN has an efficient light encoder for feature extraction and two different pixel-to-pixel mapping modules for homogeneity correction and image prior loss. With designed regional and global loss functions, our zero-shot bias correction method is able to perform online optimization with efficient execution and high performance of bias field correction. 

\subsection{Methodology}
\label{sec:method}
The bias corruption of MRI images is formulated by multiplying a bias field with some additive noise:
\begin{equation}
Y = XB + n
\end{equation}   
where Y is the observed corrupted image, X is the bias-free MRI image, B is the bias field, and n is the additive noise. Common practice of bias correction with DL is to train a model to estimate B from the training data with certain distribution. 

In our solution, we choose an alternative approach that decomposes the above problem as homogeneity refinement with image prior constraint. Specifically, as observed in MRI images with strong bias artifacts, the resulted images are encountered similar situation of low-light nature image~\cite{Zero-DCE++}, where intensities from the same region are having large variances due to imperfect detector and over/under-exposure of the information. Practically, the inhomogeneities of the dark and bright region can be iterative corrected by the function:
\begin{equation}
HC(x,y,z) = I(x,y,z) + \alpha I(x,y,z)(1 - I(x,y,z))
\end{equation}  
where HC is the homogeneity corrected output of position $(x,y,z)$ with intensity $I$ and a learned parameter $\alpha$ per data point. With high-order estimation of the HC, the pixel-based regional corruption can be adaptive refined, which is specifically defined as (without considering voxel position)
\begin{equation}
HC^n = HC^{n-1} + \alpha HC^{n-1}(1 - HC^{n-1})
\end{equation}  
where parameter $n$ is the number of iteration, which controls the strength of the homogeneity correction and was chosen to be 4 in the paper to trade-off accuracy and computation cost~\cite{Zero-DCE++}. 

The above correction would lead to the overestimation for correcting the bias corruption. To constrain the predicted $\hat{X}$ without over-exposure, the image prior loss is considered. Conventional bias correction methods directly predict the bias field $B$ with proper constraint~\cite{simko2022mri}. In contrast, without knowing the actual bias of the image, the predicted $\hat{B}$ in the proposed model is constrained by the image prior, i.e., minimizing $|Y-\hat{X}\hat{B}|$, which therefore ensures the homogeneity refined image with predicted bias field can match the input.

\subsection{CNN Model Design}
The CNN model with skip-connection design is depicted in Fig.~\ref{model_design}. The 3D volumetric MRI image is downsampled by factor of 8 before processed by Depthwise Separable Convolutions (DSC), which consists of 3D depthwise convolution with kernel size 3 and 3D pointwise convolution with kernels size 1. The purpose of these light-weight design is to reduce the online optimization cost for a standard 3D volumetric data, and accelerate the optimization. With two predicted parametric maps from the last layer (which are interpolated to original resolution), the input data with original resolution is operated by two nonparametrical modules for inhomogeneity and image prior constraint. In our design, the parametric map $\alpha$ is normalized by Tanh function, while the predicted bias map is multiplied with the predicted images to match input, normalized by Sigmoid. 
\subsection{Online Optimization}
The proposed zero-shot model performs the online optimization at testing phase. To properly constraint the model to correct bias field, the designed objective function consists of two parts for joint training.
\begin{equation}
L(\theta) = L_{hc}(\theta) + L_{prior}(\theta)
\end{equation} 

Function $L_{hc}(\theta)$ is designed to optimize the homogeneity correction function (HC) for CNN with parameters $\theta$. Specifically, this function consists of several components as below.
\begin{equation}
L_{hc}(\theta) = L_{smo}(\alpha) + L_{spa}(HC^n, Y) + L_{exp}(HC^n)
\end{equation} 
where $L_{smo}$ is the smoothness function, which is defined as
\begin{equation}
L_{smo}(\alpha) = \frac{1}{N}\sum_{n=1}^{N}\left(\sum_{k\in\{x,y,z\}}|\nabla_{k}\alpha_n|\right)^2
\end{equation} 
Loss spatial consistency loss $L_{spa}$ is defined as 
\begin{equation}
L_{spa}(HC^n, Y) = \frac{1}{K}\sum_{n=1}^{K}\sum_{j\in\Omega(i)}(|\hat{HC}_i^n-\hat{HC}_j^n| - |\hat{Y}_i-\hat{Y}_j|)^2
\end{equation} 
where $K$ is the number of local region, and $\Omega(i)$ is the 8 neighboring voxels. The symbol $\hat{*}$ denotes the average intensity value of the local volume in the image. The $L_{exp}$ is the exposure loss function to constrain the average regional values of the predicted images~\cite{Zero-DCE++}. The weights of smoothing loss are empirically selected as 1600 for $L_{hc}$, as suggested by Zero-DCE++~\cite{Zero-DCE++}.

Function $L_{prior}(\theta)$ is designed to constrain the predicted $\hat{X}\hat{B}$ to match $Y$. Specifically, this loss function is formulated as
\begin{equation}
L_{prior} = |Y-\hat{X}\hat{B}|_1 + L_{smo}(\hat{B})
\end{equation} 

\section{Experiments and Results}
\subsection{Dataset}
The proposed zero-shot bias correction method was evaluated on five different MRI datasets. First, we randomly selected 39 3D Sagittal T1w MRI brain images from ADNI dataset~\cite{langbaum2009categorical}, which are processed by ANTsPyNet~\cite{tustison2021antsx} to obtain mask of grey matter (GW) white matter (WM), and cerebrospinal fluid (CSF). Second, 59 T1w MR brain images were randomly selected from IXI dataset\footnote{https://brain-development.org/ixi-dataset/}, which were processed to have anatomy masks same as ADNI dataset. Third, 21 T2w MR brain images were randomly selected from OASIS3 dataset~\cite{lamontagne2019oasis}, the anatomy masks for GW, MW and CSF were also generated for evaluation. In addition, a private Dixon (DX) dataset with 7 patients was considered, which have water sequences (W) and fat sequences (F) for pelvis region. The sequences were segmented to have bladder (BLD), water and fat components for evaluation purposes. All the used datasets were approved by the ethical committee for research purpose. 
\subsection{Experiment settings}
During the execution stage, the designed zero-shot model is optimized by standard Adam optimizer with learning rate equals to 0.005 for ADNI/IXI/OASIS datasets, and 0.0005 for DX datasets, with weight decay equals to 0.0001. The total iteration step is selected to be 100 to balance the efficiency and accuracy.

The evaluation metric was the Coefficient of Variation (CV) for all the considered datasets via the aforementioned anatomy masks. We compared the proposed method to the state-of-the-art N4 method, which is implemented in ANTs toolbox via ITK implementation~\cite{tustison2021antsx}. 

\begin{figure}[ht!]
\centering{\includegraphics[width=7cm]{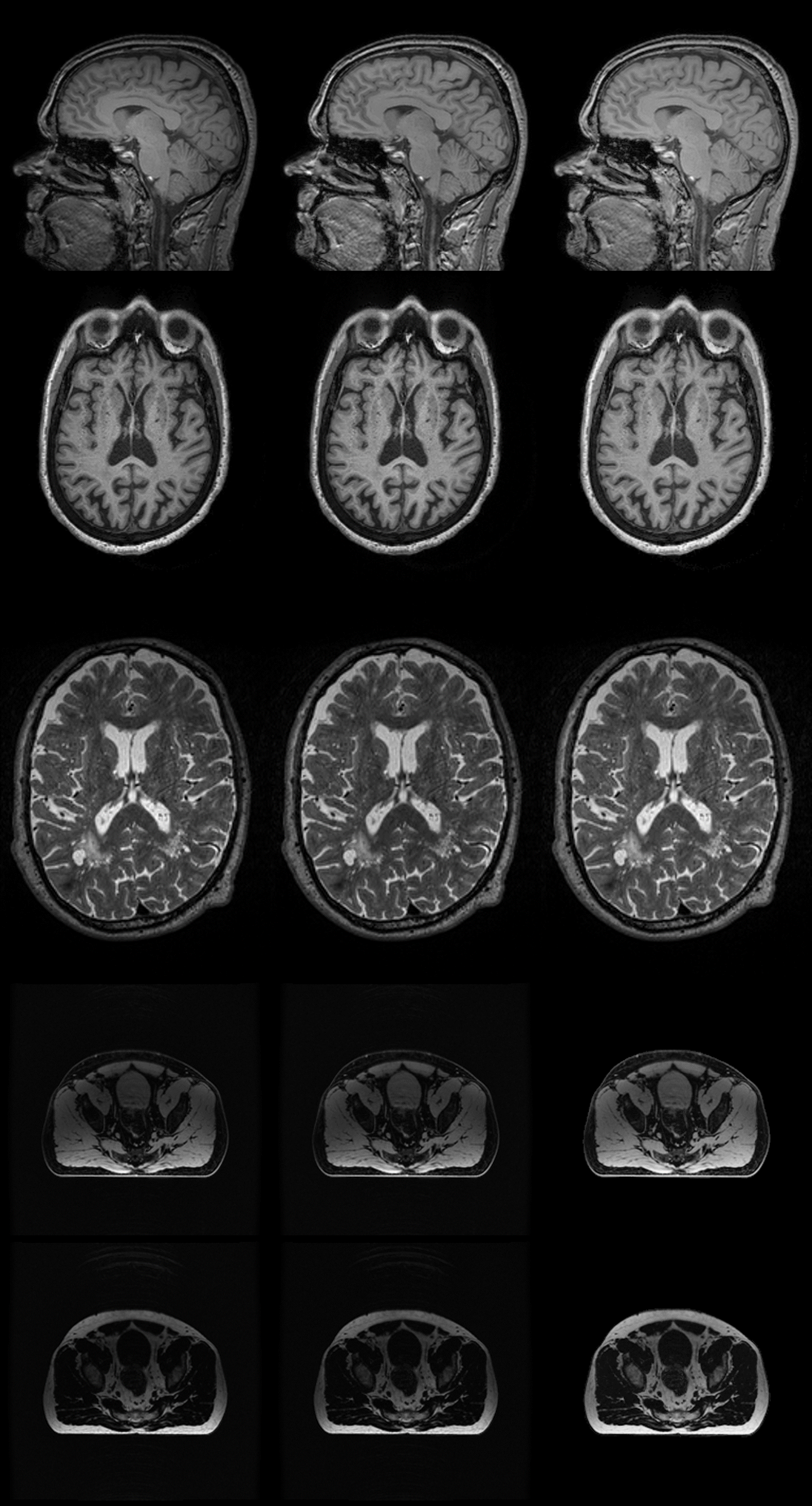}}
\caption{Example images of comparisons. From top to bottom: IXI, ADNI, OASIS, Dixon-water and Dixon-fat. First column: original image, second column: N4 result, third column: the results of the zero-shot bias correction.}
\label{img_compare}
\end{figure}

\subsection{Experiment results}
\begin{table}[tb]
\small
\centering
\caption{Evaluation results of the method on different dataset, measured by CV with mean(std.)}
\label{performance}
\begin{tabular}{lllll}
\hline
Dataset                             & Tissue & Original & N4 &  Proposed \\
\hline
\multirow{3}{*}{\makecell[l]{ADNI}} & GM     & 0.194(0.022) & 0.163(0.027) &  0.149(0.019) \\
                                    & WM     & 0.155(0.017) & 0.118(0.020) & 0.104(0.014)  \\
                                    & CSF    & 0.564(0.065) & 0.558(0.067) & 0.492(0.062)  \\
\hline
\multirow{3}{*}{\makecell[l]{IXI}}  & GM       & 0.150(0.031) & 0.120(0.013) &  0.117(0.013) \\
                                    & WM       & 0.126(0.027) & 0.084(0.007) & 0.085(0.011)  \\
                                    & CSF      & 0.586(0.075) & 0.565(0.065) &  0.519(0.057) \\
\hline
\multirow{3}{*}{\makecell[l]{OASIS}}& GM       & 0.356(0.059) & 0.338(0.064) &  0.304(0.057) \\
                                    & WM       & 0.248(0.060) & 0.234(0.064) & 0.216(0.053)  \\
                                    & CSF      & 0.694(0.083) & 0.659(0.080) &  0.619(0.070) \\
\hline
\multirow{2}{*}{\makecell[l]{DX-W}}  & BLD     & 0.159(0.036) & 0.138(0.028) & 0.118(0.029)  \\
                                     & Water       & 0.498(0.038) & 0.464(0.054) & 0.262(0.031)  \\
\hline
\multirow{2}{*}{\makecell[l]{DX-F}}  &  BLD   & 0.524(0.128) & 0.521(0.130) &  0.469(0.109) \\
                                     &  Fat       & 0.521(0.025) & 0.505(0.022) &  0.273(0.064) \\
\hline
\end{tabular}
\end{table}

The numerical performances are summarized in Table.~\ref{performance}. As can be observed, the zero-shot bias correction achieved better numerical values in Dixon dataset for both bladder and water(fat) regions. As for T1w brain images from different dataset, the proposed method achieved comparable performance to N4 algorithm on white matters, but better performance on grey matters and CSF. Example images from considered datasets are shown in Fig.~\ref{img_compare}. From the image, N4 method will fail in challenging Dixon images, while our method can better correct the inhomogenieities.  

The average execution time of the proposed method on an NVIDIA-A100 is 3 sec., while it can be degraded to around half minute on NVIDIA-P6000 for the considered ADNI dataset. In contrast, N4 method is measured on a CPU of AMD EPYC 7742 with around 800 sec. on ADNI dataset. With state-of-the-art deep learning based automated pipelines with a dedicated GPU, the proposed method shows faster processing time than the commonly used N4 algorithm, but much better results on challenging strong bias field MRIs.

\section{Conclusions}
This paper presents a novel zero-shot data-free deep learning solution for inhomogeneity correction for MR 3D volumetric data. This novel approach introduces an alternative viewpoints of solving the bias corrupted images, which showed its better ability to achieve better performance than the current solutions. With dedicated evaluation on different MR sequences, the propose achieved comparable or even better numerical results than popular N4 bias correction method, while significantly accelerating the operating speed to just seconds. Therefore, the full automated MRI pipelines, such as end-to-end automated Alzheimer's disease analysis, can be drastically accelerated for pre-processing steps. 
\section{Acknowledgement}
PREDICTOM receives funding from the Innovative Health Initiative Joint Undertaking (IHI JU), under Grant Agreement No 101132356. JU receivessupport from the European Union’s Horizon Europe research and innovation programme, COCIR, EFPIA, EuropaBio, MedTechEurope and Vaccines Europe. The UK participants are supported by UKRI Grant No 10083467 (National Institute for Health and Care Excellence), Grant No 10083181 (King's College London), and Grant No 10091560 (University of Exeter). The Swiss participant is supported by the Swiss State Secretariat for Education, Research and Innovation Ref No 113152304. See https://www.predictom.eu/ for more details
\bibliographystyle{IEEEbib}
\bibliography{strings,refs}

\end{document}